\documentclass[aps,superscriptaddress,nofootinbib]{revtex4}

\usepackage{graphicx}
\usepackage{graphics}
\usepackage{amsmath}
\usepackage{subfigure}

\begin{document}
\title{Observer Dependent Horizon Temperatures: a Coordinate-Free Formulation of Hawking Radiation as Tunneling}

\author{Sean Stotyn}
\email{smastoty@sciborg.uwaterloo.ca}
\affiliation{Department of Physics and Astronomy, University of Waterloo,\\
                   Waterloo, Ontario, Canada, N2L 3G1}
\author{Kristin Schleich}
\email{schleich@physics.ubc.ca}
\author{Donald M. Witt}
\email{donwitt@physics.ubc.ca}
\affiliation{Department of Physics and Astronomy, University of British Columbia,\\
                   Vancouver, British Columbia, Canada, V6T 1Z1}
                   \affiliation{Perimeter Institute for Theoretical Physics, \\ 31 Caroline Street North, Waterloo,
Ontario, Canada N2L 2Y5}

\begin{abstract}

We reformulate the Hamilton-Jacobi tunneling method for calculating Hawking radiation in static, spherically-symmetric spacetimes by explicitly incorporating a preferred family of frames. These frames correspond to a family of observers tied to a locally static timelike Killing vector of the spacetime. This formulation separates the role of the coordinates from the choice of vacuum and thus provides a coordinate-independent formulation of the tunneling method. In addition, it clarifies the nature of certain constants and their relation to these preferred observers in the calculation of horizon temperatures.  We first use this formalism to obtain the expected temperature for a static observer at finite radius in the Schwarzschild spacetime. We then apply this formalism to the Schwarzschild-de Sitter spacetime, where there is no static observer with 4-velocity equal to the static timelike Killing vector.  It is shown that a preferred static observer, one whose trajectory is geodesic, measures the lowest temperature from each horizon.  Furthermore, this observer measures horizon temperatures corresponding to the well-known Bousso-Hawking normalization.  

\end{abstract}

\maketitle

\section{Introduction}

A key connection in the study of black hole thermodynamics was established in 1974 when Hawking showed that black hole horizons   in Schwarzschild spacetime emit radiation in a thermal spectrum \cite{Hawking:1974sw}.  In \cite{Unruh:1976db}, Unruh demonstrated that Minkowski spacetime appears as a thermal bath to an accelerating observer, the interpretation of which is that the Rindler horizon, though not a black hole horizon, also thermally radiates.  Gibbons and Hawking then showed in \cite{Gibbons:1977mu} that, similarly, cosmological horizons also thermally radiate.  The thermal radiation emitted by any horizon has come to be known as Hawking radiation.  

Recently, a new method of computing the spectrum of Hawking radiation was proposed by Parikh and Wilczek  \cite{Parikh:1999mf}. This method treats Hawking radiation as a tunneling process and emphasizes the dynamical nature of the particle production. There are two slightly different implementations of the tunneling method; the null geodesic method of Parikh and Wilczek \cite{Parikh:1999mf} and the Hamilton-Jacobi  method of Angheben \emph{et al.} \cite{Angheben:2005rm}, which is an extension of the complex paths approach introduced in \cite{Srinivasan:1998ty}. Both methods make use of the WKB approximation, which is argued to be valid because gravitational blueshift near the horizon ensures the radiation has a wavelength much shorter than the width of the barrier.  In the WKB 
approximation, the tunneling probability, $\Gamma=\frac{P(emission)}{P(absorption)}$, where $P(x)$,
 the probability of event $x$, is related to the exponential of the imaginary part of the classical action, $S$.  This tunneling probability is then compared to a thermal Boltzmann distribution for emission, $\Gamma=e^{-\beta E}$, thereby assigning a temperature to the emitted radiation.  Thus, to derive Hawking radiation as a tunneling effect, the imaginary part of the action must be calculated.  This calculation is typically carried out in a manifestly coordinate-dependent way. This formalism reproduces various well known results  \cite{KeskiVakkuri:1996gn,Srinivasan:1998ty,Parikh:1999mf,Parikh:2002qh,Parikh:2004ih,Parikh:2004rh,Angheben:2005rm,Arzano:2005rs,Kerner:2006vu}.
However, there has been controversy over  tacit choices in various implementations of tunneling 
methods. It was noted that the tunneling formulation was not invariant under canonical transformations \cite{Chowdhury:2006sk}, and that coordinate choice issues lead to a factor of two in the calculation of the temperature of the Schwarzschild horizon \cite{Angheben:2005rm}. Various analyses and proposals for addressing these issues have been recently carried out in \cite{Angheben:2005rm,Mitra:2006qa,Akhmedova:2008dz}. However no simple, unifying picture of tunneling resolving all of these issues has yet been given. 

This paper proposes that a key step toward such a resolution is to explicitly exhibit the role of  the timelike Killing vector in the tunneling method formulation. We do so for the simple case of  tunneling  in static, spherically-symmetric spacetimes. We formulate the Hamilton-Jacobi tunneling method with respect to an orthonormal frame based on the chosen static timelike Killing vector. We show that by doing so, certain issues with coordinate invariance and canonical invariance are resolved.  An added benefit of this formalism is that it allows calculation of temperature for static observers at finite radius. Furthermore, it lends itself quite naturally to spacetimes that do not possess asymptotic regions, such as Schwarzschild-de Sitter (SdS) spacetime.

The organization of the paper is as follows:   In section 2 we formulate the tunneling calculation of Hawking radiation in the Hamilton-Jacobi formalism for static, spherically-symmetric spacetimes with explicit reference to an orthonormal frame based on the static timelike Killing vector. The resulting formulation yields a manifestly coordinate independent and canonically invariant temperature. We then show that a static observer in the spacetime measures a temperature that is scaled by the inverse length of the timelike Killing vector evaluated at the observer's radius.  
In section 3, we apply this formalism to the SdS spacetime. We compute the temperature of each horizon for a family of static observers. We show that the lowest temperature of either horizon is measured by a preferred static observer - one whose trajectory is geodesic - and that the temperatures measured by this preferred geodesic observer correspond to the well-known Bousso-Hawking normalization.  We conclude with a discussion of these results and possibilities for their generalization.

\section{Coordinate Independent formulation of Tunneling}
\setcounter{equation}{0}

In quantum field theory in curved spacetime (See, for example, \cite{Birrell:1982ix}), the quantization of a free scalar field involves a) solving the classical scalar field equation to find a complete set of orthonormal states, b) a division of this set into two subsets; those of positive energy and those of negative energy, and c) the construction of the quantum field operator by promotion of the Fourier coefficients of the complete set to creation and annhilation operators, the choice being dependent on the sign of the energy. When the background is Minkowski spacetime, the split into positive and negative energy modes is usually made with reference to a global static Killing vector (for example $t^\mu = \partial_t$ in the coordinates $ds^2 = -dt^2 + dx^2 + dy^2 + dz^2$). This split is Lorentz-invariant; consequently there is a unique Lorentz-invariant Minkowski vacuum. In curved spacetime, or for choice of static Killing vector that is not global (such as that given by $t^\mu 
 = \partial_t$ in $ds^2 = -a^2x^2 dt^2 + dx^2$  in Rindler coordinates for two dimensional Minkowski spacetime) the split into positive and negative energy sets is not preserved by the natural classical symmetry of curved spacetime, diffeomorphism invariance. Consequently, and famously, different choices for the timelike  vector used to define the split between positive and negative energy modes may lead to inequivalent vacua. One of these inequivalent vacua may be a thermal state, or exhibit infinite particle production when this state is exhibited in terms of an alternate set of modes corresponding to a different vacuum.  Hence specification of the timelike vector used to define the split of modes is key to the definition of the vacuum and its properties as seen by its observers.

In contrast, in the calculation of temperature through tunneling methods, no explicit identification of positivity of energy is  made. Instead, the identification of the positivity of energy is implicit in the coordinatization and anzatz used for the calculation of the imaginary part of the action.  However, we can remove this implicit identification by carrying out the analysis through an explicit introduction of the preferred orthonormal frame associated with the static Killing vector.  For clarity, we will carry out this construction for the simple case of a static spherically-symmetric spacetime in the Hamiton-Jacobi formalism.
 
Consider a general spherically-symmetric, static spacetime with a locally static timelike Killing vector, $k^\mu$ with norm $k^\mu k_\mu = -f$, and spatial Killing vectors, $\xi^\mu_i$, $i = 1,2,3$ which are generators of an $SO(3)$ algebra, which commute with $k^\mu$ and which act freely on the spacetime whose orbits are spatial spheres.  A family of static observers whose 4-velocities are proportional to the timelike Killing vector is given by
\begin{equation}
e_t^\mu = \frac 1{\sqrt{f}}k^\mu.
\end{equation}
This family can be taken as the timelike vector field of  a spherically-symmetric  orthonormal frame,
$e^\mu_r$, $e^\mu_j$, $j= 1,2$ chosen such that ${\mathcal L}_k e_r = {\mathcal L}_k e_j = 0$ and ${\mathcal L}_{\xi_i} e_r = 0$, where $\mathcal L$ denotes the Lie derivative.  
The metric for a static spherically-symmetric spacetime can be expressed in terms of this orthonormal frame as
\begin{equation}
g^{\mu\nu} = \eta^{ab} e^\mu_a e^\nu_b
\end{equation}
where $\eta^{ab}$ is the Minkowski metric. An example of such an orthonormal frame for a commonly chosen coordinate chart given in the Appendix. The derivation that follows, however, does not rely on this chart.

The Killing vector field $k^\mu$ is timelike in the region where $f>0$. The {\it Killing horizon} is a connected set of points where $f$ vanishes  that form  a null hypersurface. The Killing horizon and the black hole horizon coincide in the Schwarzschild spacetime. However, Killing horizons need not be black hole horizons: the Killing horizon of the de Sitter spacetime and the Killing horizons of the Nariai spacetime are cosmological horizons.

We now formulate the Hamilton-Jacobi tunneling method of \cite{Angheben:2005rm} by explicit inclusion of the above family of orthonormal frames.  The classical action, $S$, is taken to satisfy the relativistic Hamilton-Jacobi equation
$g^{\mu\nu}\partial_{\mu}S\partial_{\nu}S+m^2=0$.
Writing $g^{\mu\nu}$ in terms of the orthonormal frame yields
\begin{equation}
\eta^{ab}e_a^{\mu}\partial_{\mu}Se^\nu_b\partial_{\nu}S+m^2=0. \label{eq:HJ2}
\end{equation}
We now make the anzatz that the dominant contribution to the amplitude is the s-wave, or most symmetric one; consequently, 
$S$ must be invariant under rotations: $\mathcal L_{\xi_i}S =  \xi^\mu_i\partial_\mu S= 0$. It follows that $e^\nu_j\partial_{\nu}S = 0$ for the angular basis vectors.  As there is a further symmetry generated by $k^\mu$, $S$ must be covariantly constant under its action: ${\mathcal L}_k S= k^\mu\partial_\mu S = -\alpha$, where $\alpha$ is a constant.  The physical interpretation of this is that $\partial_\mu S$ is cotangent to a geodesic and  $\alpha$ is a conserved quantity along that geodesic.\footnote{This constant is often identified as the energy (see, for example \cite{Kerner:2007rr}). However, as discussed later, it is in fact, in general the energy  seen only by a certain preferred observer.}
Because $e_t^\mu = \frac{1}{\sqrt{f}} k^\mu$,  (\ref{eq:HJ2}) becomes simply
\begin{equation}
-\frac{\alpha^2}{f} + (e_r^{\mu}\partial_{\mu}S)^2+m^2=0. \label{eq:HJ3}
\end{equation}
Now, $e_r^\mu \partial_\mu = \frac {\partial}{\partial \sigma}$ is the  derivative with respect to the proper distance and therefore
\begin{equation}
S = \pm\int d\sigma\sqrt{ \frac{\alpha^2}{{f}}-m^2} + \Gamma \label{eq:HJ4}
\end{equation} 
where $\pm$ corresponds to outgoing (+) and ingoing (-) particles and $\Gamma$ is a function of integration independent of $\sigma$.  Furthermore, since $k^\mu\partial_\mu = \frac {\partial}{\partial \lambda}$ is the derivative with respect to the the parameterization coordinate of the Killing vector, $\lambda$, and $k^\mu\partial_\mu S=-\alpha$,
\begin{equation}
S =-\alpha\lambda+ \Gamma'\label{eq:HJ5}
\end{equation}
where $\Gamma'$ is a function of integration independent of $\lambda$. Putting these together for the case of massless particles, $m^2 = 0$, yields 
\begin{align}
S = -\alpha \lambda \pm W + C\label{eq:HJ6.0} \\
W=\int \frac{\alpha}{\sqrt{f}}d\sigma \label{eq:HJ6}
\end{align} 
where $C$ is an arbitrary constant of integration whose significance will be explained shortly. An imaginary part of (\ref{eq:HJ6.0}) comes from either  a pole in the integral over proper distance (\ref{eq:HJ6}) and/or from a complex part of the constant $C$. The presence of a pole corresponds to a Killing horizon for the static Killing vector.

The expression for the action in (\ref{eq:HJ6.0}) is very similar to the standard anzatz,  $S=-Et+W(r)+C$, for s-wave solutions in the Hamilton-Jacobi formalism but there are a few important key differences. 

 First, 
as (\ref{eq:HJ6}) is expressed as an integral over the proper distance, it is manifestly coordinate invariant.  In the standard formulation, the Hamilton-Jacobi method introduced the proper spatial distance after the related term was first formulated in a coordinate chart as a means to circumvent certain problems in computation \cite{Angheben:2005rm}.  Here, the integration over proper distance occurs naturally through the formulation in terms of the static orthonormal frame.  

Second, the constant $\alpha$ is no longer identified with the energy of the null s-wave; rather it is a constant of the motion.  In order to identify it with an energy, one needs to utilize the orthonormal frame associated with an observer who will see that energy.  In particular, the energy seen by a static observer at spatial point $0$ is given by
\begin{equation}
E_0 = -e_t^\mu \partial_\mu S = \frac \alpha{\sqrt{f_0}} \label{eq:energy}
\end{equation}
where $f_0$ is evaluated at the position of the static observer.  Clearly, in Schwarzschild spacetime, the energy seen by a static observer at infinity, $E_\infty$, is identical to $\alpha$ because $f_\infty = 1$.  In previous implementations of  tunneling methods, this identification has been implicitly carried out; the constant $E$ used in the usual anzatz  is the energy seen by a static observer at infinity, $E_\infty$. Application of these techniques in de Sitter spacetime also implicitly choose this constant to be related to a single preferred observer, in this case one at the origin of the static chart, where $f_0=1$. However, for more general spacetimes there may be no position $p$ at which $f_p=1$. Therefore,  is not clear what the application of the standard formula for temperature obtained by tunneling methods means in such spacetimes as it is not clear where that temperature is measured.  In contrast, our formulation makes precise the relation between the constant $\alpha$ and the energy seen by a static observer at any position.  Therefore it can be applied in any static spacetime to obtain a physically meaningful result for any static observer.

Third, our derivation clearly demonstrates that $\lambda$ is an everywhere real parameter that parameterizes integral curves of the Killing vector $k^\mu$. This parameter, of course, remains real regardless of whether the Killing vector is timelike, spacelike or null.  It follows that in the Hamilton-Jacobi anzatz the parameter $t$  is also an everywhere real parameter associated with the timelike Killing vector used in the formulation. This point has led to some confusion in the literature.  It was stated in \cite{Akhmedova:2008au,Akhmedov:2008ru,Banerjee:2008cf} that there is an imaginary contribution to the action from the $-Et$ contribution to the action because $t$ and $r$ change their timelike and spacelike nature as the horizon is crossed.  Our derivation shows that this is clearly not the case.  It is therefore evident that the \emph{only} imaginary contribution to the action can come from $W$ and $C$ in (\ref{eq:HJ6.0}).  

Finally, it is now manifest that the existence of an imaginary contribution to the action is associated with an integral through a \emph{Killing horizon}: if there exists a simple pole in the integrand of  (\ref{eq:HJ6}) then the action picks up an imaginary part. Since it is the Killing horizon that is of interest, the imaginary contribution to the action is, of course, dependent upon the choice of static Killing vector and hence so is the horizon temperature.  This makes explicit where the choice of vacuum enters in the tunneling formalism.  Inequivalent static timelike Killing vectors correspond to potentially inequivalent vacua so the temperature calculated depends on the choice of the Killing vector.  This formalism is, however, independent of the coordinate chart that the Killing vector is expressed in. This identification is implicitly carried out in the use of the static chart in the Hamilton-Jacobi method \cite{Angheben:2005rm}; our formulation shows that the key factor is the norm of the Killing vector, not use of the static chart.
The integration over proper distance (\ref{eq:HJ6}) can  be implemented in different coordinate charts; however, all charts will yield the same final answer as illustrated in the Appendix.

At this point, the remaining issue is a proper evaluation of the integral  (\ref{eq:HJ6}) and its use in the calculation of the temperature. We begin by reviewing the discussion of the issues involved in this as discussed previously in the literature.

 Since the timelike Killing vector has vanishing norm on the horizon, both the ingoing and outgoing particles have a pole at the horizon that needs to be accounted for. Therefore $W$ will have an imaginary contribution for both ingoing and outgoing particles. If one requires that the probability of absorption be unity in the classical limit,  then, as pointed out by Mitra in \cite{Mitra:2006qa},  the constant of integration, $C$, should be allowed to be complex and its value chosen to ensure this. This requirement fixes  $\mathrm {Im} C=-\mathrm {Im}W_-$  where $W_-$ is the amplitude for ingoing particles.  Now, in any static chart the imaginary part of the action for an ingoing particle is related to that of an outgoing particle, $W_+$,  by $\mathrm{Im} W_-=-\mathrm{Im} W_+$.  It then follows that the imaginary part of the action for outgoing particles is given by $\textrm {Im}S=\textrm {Im}(W_++C)=2\textrm{Im}W_+$.

Using  Schwarzschild coordinates to evaluate $W_+$ produces a temperature that is twice the Hawking temperature \cite{Angheben:2005rm} . This factor of two in the temperature corresponds to a factor of one half in the imaginary part of the action.  However,  the previous argument implies that the imaginary part of the action for outgoing particles is actually equal to  $2\textrm{Im}W_+$, not $W_+$. Hence the factor of one half is cancelled by the factor of two included by properly accounting for ingoing particles. In summary,  the standard Hawking temperature is  recovered by taking detailed balance into consideration.  

However it was alternately proposed in \cite{Angheben:2005rm} that to remove this factor of two from Schwarzschild coordinates, one should use a coordinate chart regular on the horizon, the natural choice involving the proper distance.  In doing so, the authors were able to reproduce the standard Hawking temperature without including the factor of two from detailed balance. However, since the coordinate chart used in \cite{Angheben:2005rm}  is also static, detailed balance should apply in this case as well. Doing so will introduce a factor of one half in the temperature.  Thus there is now, naively, the opposite discrepancy: A calculation in Schwarzschild coordinates yields the proper Hawking temperature whereas one in proper spatial distance seemingly yields \emph{half} the Hawking temperature if detailed balance is taken into account.  

We have argued, however, that our formalism is independent of the coordinate chart used to express the timelike Killing vector. Indeed, (\ref{eq:HJ6}) is formulated in terms of proper distance alone. Hence it should yield the Hawking temperature. Moreover, when its evaluation is carried out in different coordinate charts, all should yield the same temperature as well.  This is indeed the case; the proper choice of contour in the integral over proper distance for the detailed balance approach yields the desired result.
\begin{figure}
    \centering
     \subfigure[]{\includegraphics[width=3.2 in]{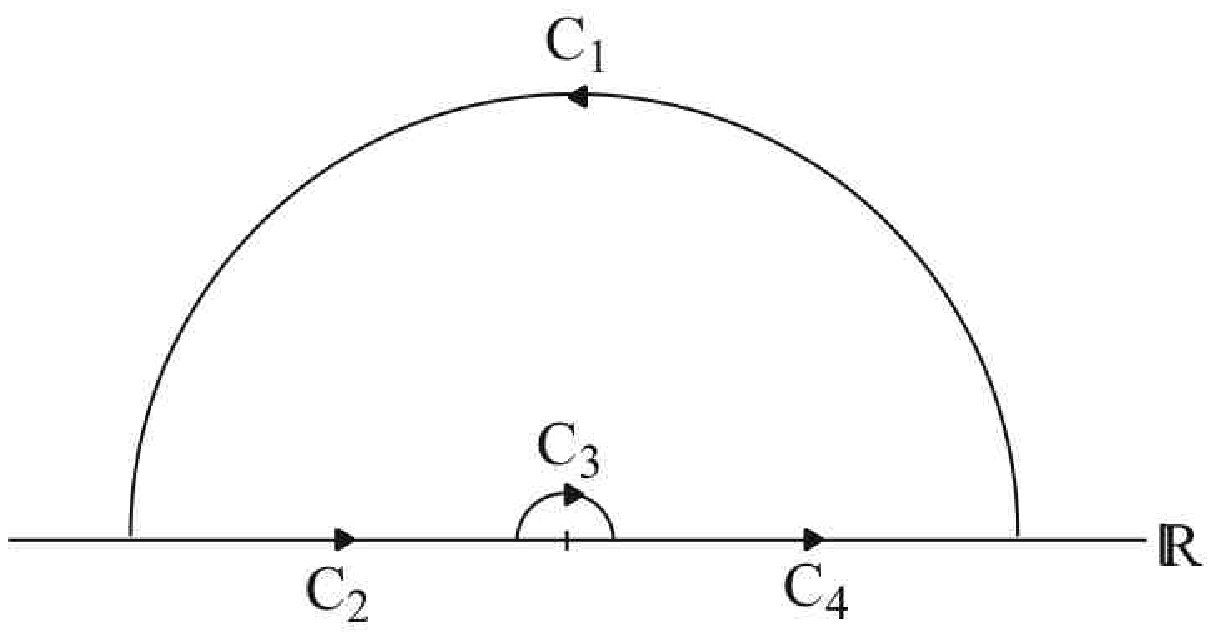}}
     \hspace{.3 in}
     \subfigure[]{\includegraphics[width=2.25 in]{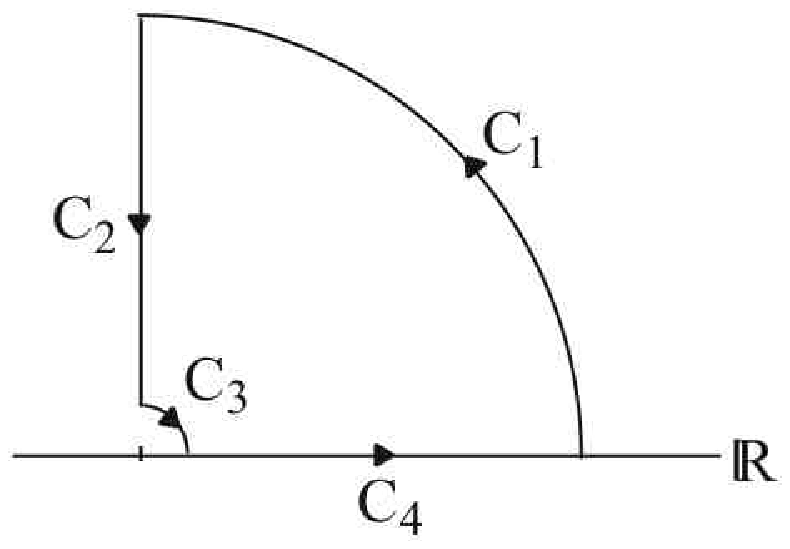}}
     \caption{(a) A common contour erroneously used in the Hamilton-Jacobi tunneling method when working with the proper spatial distance.  This is the proper contour to use, however, if one is working in Schwarzschild coordinates.  (b) The proper contour to use when working with the proper spatial distance.  In both figures a) and b), the tick mark on the real axis denotes the position of the Killing horizon.}
     \label{fig:contour}
\end{figure}

To see this, note that for a static spacetime, if $f<0$ inside the Killing horizon then $e_t^\mu$ is a spacelike vector and $e_r^\mu$ is a timelike vector inside the Killing horizon. This implies that
the infinitesimal proper distance $d\sigma$, thought of as the square root of $e_r^\mu {e_r}_\mu$, and consequently the proper distance itself are both imaginary inside the horizon. In order to capture the notion of tunneling from inside the Killing horizon to outside, the lower bound in terms of $\sigma$ must be imaginary; of course the upper bound will remain real. Hence we argue that one ought to use the quarter-contour shown in Fig.~\ref{fig:contour}b.  In our coordinate-invariant construction, the quarter-circular contour is essential and arises naturally from the formulation of the integral in terms of proper distance.  The imaginary contribution to the action comes from jumping around the pole, hence the quarter-circular contour yields an imaginary contribution which is half that obtained using the semi-circular contour. Hence the same factor of one half that appeared when using Schwarzschild coordinates is seen to appear when using proper distance.  This factor of one half will cancel the factor of two when taking detailed balance into account.

The temperature now follows by the standard identification. If the static spacetime has a simple pole at the Killing horizon,  then the imaginary part of the integral in (\ref{eq:HJ6}), using the proper contour, results in $\mathrm{Im} W_+ = -\frac{\pi}{2} \alpha P$ where $P$ is a constant depending on the details of the spacetime. Hence, since $2\mathrm{Im}S = 4\mathrm{Im} W_+$, one finds $2\mathrm{Im}S=-2\pi \alpha P$.  Now note that a thermal distribution as seen by a static observer at a spacetime point $0$ is given by $e^{-\beta_0 E_0}$, where $E_0$ is the energy of the radiation as measured by the observer.  The tunneling probability at this point is similarly given by $e^{2\mathrm{Im}S}$. Connecting the tunneling probability with the thermal emission distribution gives $\beta_0 E_0=2\pi\alpha P$.  Next, recall that (\ref{eq:energy}) yields $E_0=\alpha/\sqrt{f_0}$. Thus
\begin{equation}
\beta_0= 2\pi P\sqrt{f_0} \ .
\end{equation} 

For asymptotically flat spacetimes, the inverse temperature at asymptotic infinity, where $f=1$, is given by  $\beta_H=2\pi P$, which is the inverse of the standard Hawking temperature. Explicit calculation for Schwarzschild spacetime yields $P=\frac1\kappa$, where $\kappa= \frac 1{4M}$ is the horizon surface gravity, so indeed the Hawking result $\beta_H=\frac{2\pi}{\kappa}$ is recovered. For this case, 
\begin{equation}
\beta_0= \sqrt{f_0} \beta_H\ .
\end{equation} 
Hence the temperature, $T_0=\beta_0^{-1}$, as seen by a static observer scales by a factor inversely proportional to the norm of the static timelike Killing vector evaluated at the observer's position; this is the expected result of Tolman \cite{Tolman:1934}. Thus our derivation yields the Hawking temperature for an observer at infinity in an asymptotically flat spacetime as well as the more general formula for the temperature seen by static observers at any position.

Finally, note that $2\pi P=\frac 1\alpha W$ where $W$ is integrated on a closed contour around the pole at the Killing horizon. Consequently, the temperature is manifestly invariant under canonical transformations as it should be as noted in \cite{Chowdhury:2006sk}.

Although the above calculation is given in terms of proper distance, if desired, one can then make an appropriate coordinate transformation and correspondingly change the contour to suit the new chart.  This change in contour with change in coordinates has been previously noted; for instance,  in \cite{Akhmedov:2006pg,Akhmedov:2006un}, isotropic coordinates are used with a quarter-circular contour while Schwarzschild coordinates are used with a semi-circular contour Fig.~\ref{fig:contour}a.  A related comparison of various coordinate charts for 2D stringy black holes is given in \cite{Vagenas:2001qw}.
Near the horizon, the isotropic radial coordinate, $\rho$, and the Schwarzschild radial coordinate, $r$, are related to the proper distance via $(\rho-\rho_h)\propto\sigma$ and $(r-r_h)\propto\sigma^2$, where $\rho_h$ and $r_h$ denote the position of the horizon in the respective coordinate charts; it is the respective linear and quadratic dependences on $\sigma$ in these two charts that dictates which contour must be used.  

Finally, note that in the literature, for example in \cite{Kerner:2006vu,Angheben:2005rm},  the integral $\int{\frac{d\sigma}{\sigma}}$ is sometimes evaluated using the contour shown in Fig.~\ref{fig:contour}a, or an equivalent variation thereof.  However, if this contour is used for integration in proper distance directly, it integrates from real negative $\sigma$ to real positive $\sigma$ around the pole; a factor of two is erroneously included. 
 
We conclude with a  simple application of this formalism, to flat spacetime in two dimensions. For this case, the static chart can be given in the form
\begin{equation} ds^2 = - f(x) dt^2 + \frac 1{g(x)} dx^2\ .\label{rmetric} \end{equation}
If one chooses the global timelike Killing vector for Minkowski spacetime, which can be coordinatized in the form (\ref{rmetric}) with $f(x) = g(x) = 1$, one sees that (\ref{eq:HJ6}) is purely real. Thus, as expected, there is no Hawking radiation as there is no Killing horizon and the vacuum is the Lorentz-invariant Minkowski vacuum.  In contrast,  the choice of a locally static timelike Killing vector physically associated with accelerated observers in flat spacetime can be given in the form of (\ref{rmetric}) with $f(x) = a^2x^2$, $g(x) = 1$ where $a$ is the observer's acceleration. This choice of Killing vector yields a Killing horizon at $x = 0$ and hence (\ref{eq:HJ6}) is complex.  Direct calculation, for example by using the formulas given in the Appendix, shows the temperature of this horizon is given by $T=\frac a{2\pi}$, which is the standard Unruh temperature\footnote{One might be worried about the normalization factor of ${f_0}^{-1/2}$ 
  introduced in our construction. Recall that the Rindler observer is at a position $x_0=a^{-1}$ away from the Rindler horizon.  Therefore, $f_0=a^2(a^{-1})^2=1$ so the normalization factor, although still present, is trivially unity.}.  A different coordinatization of the same locally static timelike Killing vector is given by (\ref{rmetric}) with  $f(x) = 1-a^2x^2$, $g(x) = \frac {1-a^2x^2}{a^2x^2}$.  In these coordinates, the observer is located at $x=0$ and there is a Rindler horizon at $x_o=\pm a^{-1}$.  Evaluation yields $T=\frac a{2\pi}$ for the observer at this position, again agreeing with the Unruh temperature, as it ought to.  This example demonstrates explicitly that different choices of the timelike Killing vector correspond to different choices of the vacuum, while different parameterizations of the same Killing vector yield the same vacuum: choosing
  the global Killing vector yields an observer who sees an empty vacuum whereas choosing the locally static Killing vector yields one who sees a thermal bath.

\section{Temperature in Schwarzschild-de Sitter}

A convenient calculation of the temperature for static spacetimes is given by its identification with the periodicity of the time coordinate needed to construct a smooth instanton by Euclideanization\cite{Hartle:1976tp}.  Euclidean methods can be applied to spacetimes with positive or negative cosmological constant, however a straightforward application of these methods fails for spacetimes with multiple horizons such as the Schwarzchild-de Sitter  and Reissner-Nordstr\"om-de Sitter solutions. In general, the Euclideanized solutions corresponding to these spacetimes can not be made regular by a single choice of periodicity for the time coordinate as regularity at each horizon requires a different periodicity.\footnote{Exceptions to this circumstance occur for certain parameter values; for example the lukewarm ( $Q^2=M^2$)  Reissner-Nordstr\"om-de Sitter black hole has a regular Euclideanization \cite{Brill:1995hb}.} The interpretation of this fact is that the horizons are at
  different temperatures.  This 
 interpretation is independently supported by quantum field theory calculations, which find no choice of vacuum that removes divergences on both horizons \cite{Hiscock:1989yw,Romans:1991nq}.  
 
 The treatment of Hawking radiation as a local tunnelling process  provides an alternative to Euclidean and field theoretic computations of temperature.  It is therefore interesting to see if its application to the case of SdS sheds light on this meaning of temperature in this spacetime.
 
The analog of the Schwarzschild solution in the presence of a positive cosmological constant is the Schwarzschild-de Sitter spacetime.  Its line element in the static chart  can be written as
\begin{equation}
ds^2=-f(r)dt^2+\frac{1}{f(r)}dr^2+r^2d\Omega^2 \label{eq:SdS}
\end{equation}
where
\begin{equation}
f(r)=1-\frac{2M}{r}-\frac{\Lambda}{3}r^2\ , \label{eq:SchGauge}
\end{equation}
$d\Omega^2$ is the round metric on the unit 2-sphere, $M$ is interpreted as the mass of the black hole and $\Lambda$ is the cosmological constant.

There exist three real roots of $f(r)=0$, two positive and one negative, provided $\Lambda$ takes values in the range $0<\Lambda<\frac{1}{9M^2}$;
\begin{align*}
r_1&=-6MlR\\
r_h&=3MlR\left(1-\sqrt{1-\frac{1}{lR^3}}\right)\\
r_c&=3MlR\left(1+\sqrt{1-\frac{1}{lR^3}}\right)
\end{align*}
where $l = (9M^2\Lambda)^{-1/2}$ and $R=\cos\left(\frac{1}{3}\cos^{-1}\left(\frac{1}{l}\right)\right)$ evaluated using the smallest  angular value of $\cos^{-1}$. Note $l\ge 1$ for $0<\Lambda<\frac{1}{9M^2}$.  The two positive roots determine the positions of the black hole horizon $r_h$ and cosmological horizon $r_c$; $r_1$ is always a real root of the cubic no matter the range of $\Lambda$ but has no physical significance.

Clearly the SdS solution is static and the Killing vector has length
\begin{equation}
|k|=\sqrt{1-\frac{2M}{r}-\frac{\Lambda}{3}{r}^2}\ . \label{eq:timeKV}
\end{equation}
Hence, an application of the results of the last section show that an observer at radius $r$ situated in the static region between the two horizons sees the temperatures
\begin{align}
T_h&=\left(\frac{1}{\sqrt{1-\frac{2M}{r}-\frac{\Lambda}{3}{r}^2}}\right)\left[\frac{1}{4\pi}\left(\frac{2M}{{r_h}^2}-\frac{2\Lambda r_h}{3}\right)\right]\nonumber\\
T_c&=\left(\frac{1}{\sqrt{1-\frac{2M}{r}-\frac{\Lambda}{3}{r}^2}}\right)\left[\frac{1}{4\pi}\left(\frac{2\Lambda r_c}{3}-\frac{2M}{r_c^2}\right)\right]
\end{align}
where subscripts $h$ and $c$ denote the black hole and cosmological horizons respectively.  The cosmological horizon temperature has a minus sign with respect to the black hole temperature because the null waves in that case are moving inward; the temperature is therefore still positive.  The temperature of each horizon as seen by a static observer in this region varies with position. 

There is a preferred observer in the SdS spacetime at a particular radius, $r_o$,  where the gravitational attraction to the black hole is exactly balanced by the cosmological inflation.  To find this observer,  note that the effective potential, $f(r)=1-\frac{2M}{r}-\frac{\Lambda}{3}r^2$, has a maximum between the two horizons. Hence there is a geodesic at this maximum, whose 4-velocity is proportional to the static Killing vector.  The energy per unit rest mass of the observer at this critical radius is given by $\epsilon_o=\sqrt{1-(9M^2\Lambda)^{1/3}}$ so if an observer is stationary at $r_o$ with an initial energy per unit rest mass of  $\epsilon_o$, then that observer will stay stationary at this radius.
This observer measures the lowest temperature out of all other static observers.  
The lowest temperature is at the point where $\frac d{dr} |k|^{-1}=0$; this point is again $r_o$. Thus
we define the temperature of the horizons for SdS spacetime to be\footnote{For a  discussion of the key points in this section for 2+1 and 3+1 SdS solutions from an alternate perspective, see \cite {Corichi:2003kd} and for higher dimensional SdS solutions, see  \cite{Wu:2008rb}.}
\begin{align}
T_h&=\left(\frac{1}{\sqrt{1-(9M^2\Lambda)^{1/3}}}\right)\left[\frac{1}{4\pi}\left(\frac{2M}{{r_h}^2}-\frac{2\Lambda r_h}{3}\right)\right]\nonumber \\
T_c&=\left(\frac{1}{\sqrt{1-(9M^2\Lambda)^{1/3}}}\right)\left[\frac{1}{4\pi}\left(\frac{2\Lambda r_c}{3}-\frac{2M}{r_c^2}\right)\right]\ .
\end{align}
These normalized temperatures exactly correspond to the Bousso-Hawking normalization \cite{Bousso:1996au}. Hence this formulation of tunneling produces a direct connection between these normalized temperatures and the observers who measure them.

This result is in contrast to that of \cite{Shankaranarayanan:2003ya} which, although the Bousso-Hawking normalization was used, concludes that both horizons are at the same temperature.  It was stated in \cite{Shankaranarayanan:2003ya} that in a multiple horizon spacetime, an effective temperature can be defined by means of defining an effective horizon.  This idea of an effective horizon, however, leads to logical complications. The motivation for the effective horizon comes from a complex path approach in which the path is chosen to start inside the black hole horizon and end outside the cosmological horizon.  In this way, there is a contribution to the imaginary part of the action from both horizons, which is proportional to the inverse harmonic sum of the horizon surface gravities.   The approach employed in \cite{Shankaranarayanan:2003ya} yields the same temperature for all horizons because all temperatures contain the same inverse harmonic sum of all other horizon surface gravities.
Now consider a hypothetical spacetime comprised of a number of concentric horizons with an observer stationed inbetween two of them.  With this picture, if the temperature of the spacetime is defined via an inverse harmonic sum of the surface gravities of all horizons present, one would have direct access to information beyond a horizon just by measuring a temperature, \emph{i.e.} they would be able to make conclusions about the structure of spacetime in regions to which they do not have access.  Instead, we argue the observer would only be able to measure the temperatures of the two horizons that bound their region of spacetime.   In particular, the problem with this implementation of the complex paths approach is that one ought to take the path from inside each the horizon out to the observer, \emph{i.e.} evaluate along two separate paths, one for each horizon in order to compute each temperature.

 Now, if a spacetime has multiple horizons and a thermometer is placed somewhere in the spacetime, one should be able to read off a single temperature and so the idea of defining an effective temperature is well-warranted, albeit a little subtle.  The temperature of the radiation is defined by the emitted radiation forming a black-body spectrum.  If blackbody $A$ is at temperature $T_A$ and blackbody $B$ is at temperature $T_B$, then the spectrum obtained from combining the radiation emitted by both $A$ and $B$ will form the sum of the two blackbody spectra of $A$ and $B$.  This combined spectrum generically will not resemble another blackbody spectrum at some intermediate temperature $T'$.  Thus, in a multiple horizon spacetime, the radiation that an observer measures will simply be the sum of the various thermal spectra of the observable horizons, so one must ask what temperature a thermometer would measure if it were subject to multiple blackbody spectra at once.

\section{Summary and Discussion}

We formulated the Hamilton-Jacobi tunneling method in a coordinate-free form via constructing a family of static orthonormal frames determined by the choice of a static timelike Killing vector.  Using this formulation we were able to elucidate the role of the timelike Killing vector and its connection to the temperature.  This allowed a derivation of the horizon temperatures in a manifestly coordinate-invariant form.  This construction illuminates the place where the vacuum is chosen in the tunneling methods: the energy of the null waves is linked to the static timelike Killing vector and it is in choosing this vector that corresponds to the distinction between positive and negative energy modes in the field theory approach.  This connection to the timelike Killing vector was previously obscured because all calculations were done in a coordinate dependent form.

Furthermore, this formulation allows the direct calculation of horizon temperatures as seen by any static observer.   Previous formulations of the tunneling method, however, usually either implicitly assume an observer in an asymptotic region where the timelike Killing vector has unit length or compute a temperature in a spacetime without explicit identification of the observer who would detect it. The formalism presented herein makes no such assumption and lends itself quite naturally to spacetimes that do not contain asymptotic  regions. 
 As a direct application of our formalism, we have shown that different choices for the timelike Killing vector in two-dimensional flat space gives rise to physically distinct vacua, one being the Lorentz-invariant Minkowski vacuum measured by an inertial observer and the other being the Unruh thermal bath measured by an accelerated observer. 
 
 Application of the tunneling formalism to the Schwarzschild-de Sitter solution yields  black hole and cosmological horizons at different temperatures without the issues in interpreting a Euclidean instanton whose conical singularities from both horizons cannot be smoothed simultaneously.   Instead of an asymptotic observer, there exists a geodesic observer at constant radius who measures the lowest temperature from each horizon than any other static observer.  This static, geodesic observer was then used as the reference observer in SdS and the properly normalized temperature measured by this reference observer was shown to correspond to the Bousso-Hawking normalization.

The primary utility in treating Hawking radiation as a quantum tunneling phenomenon comes from its treatment as a local process - indeed this is the nature of its ability to avoid certain issues that plague the Euclideanization method.  Previous formulations of Hawking radiation as quantum tunneling, however, still had their shortcomings.  It was unclear that there is an important and explicit role played by the timelike Killing vector and it was not obvious why the tunneling formulation is not invariant under coordinate transformations.  The formulation of the Hamilton-Jacobi tunneling method in terms of a family of static, orthonormal frames explicitly tied to a preferred, static, timelike Killing vector, as presented in this paper, is a step in the direction to solving these shortcomings.  It would prove beneficial to generalize our results to stationary and axially-symmetric spacetimes to further elucidate the roles of specific symmetries in the tunneling formalism.  

Very recently, while this paper was being prepared,
 Cai \emph{et al.} pointed out the tie between an observer and the Kodama vector  \cite{Kodama:1979vn} in the context of identifying the apparent horizon leading to a thermal temperature in a FRW universe \cite{Cai:2008gw}. This interesting paper
 used the Kodama vector to identify the energy used in the tunnelling formalism for the temperature calculation, similar to our use of the Killing vector in the calculation of the action. However, the rest of the calculation was carried out with reference to a particular coordinate chart and makes the usual implicit identification of the preferred observer. It would clearly be interesting to extend our construction to this case as well.
 
\bigskip

\section*{Acknowledgements}

This work was supported by the Natural Sciences and Engineering Research Council of Canada. KS and DW would like to thank the Perimeter Institute for its hospitality during the writing of this paper. We would also like to thank Robert Mann for helpful discussions.

\section*{Appendix}

The line element of a static, spherically symmetric metric is often written in a coordinate chart that explicitly exhibits these symmetries
\begin{equation}
ds^2 = - f(r) dt^2 + \frac{dr^2}{g(r)}+h(r)d\Omega^2 \label{eq:gen_metric}
\end{equation}
where $d\Omega^2$ is the round metric on a unit two sphere.\footnote{This form of metric was introduced in a related context by \cite{Vagenas:2002hs}. The functions $f(r)$ and $g(r)$ are equal for many well known solutions such as the Schwarzschild and Reissner-Nordstr\"om spacetimes, but they need not be for more general theories of gravity and more general solutions.  } The orthonormal frame associated with this coordinate chart can be explicitly written:
$e_t^\mu =  \frac 1{\sqrt{f}} k^\mu$, $ e_r^\mu = \sqrt{g} r^\mu$,  $e_j^\mu = \frac {1}{\sqrt {h(r)}}\theta_j^\mu$
where $k^\mu = \partial_t$, $r^\mu = \partial_r$ are the coordinate basis vectors associated with $r$ and $t$
 and $\theta_i^\mu$ are a set of orthonormal basis vectors on the unit sphere.

The integration over proper distance (\ref{eq:HJ6}) can also be explicitly implemented. The norm $f(r)$  and proper distance are implicit functions of the coordinate $r$. In a neighborhood of a Killing horizon at $r_0$, one has
\begin{eqnarray}
&f(r)=f'(r_0) (r-r_0)+  \frac 12 f''(r_0) (r-r_0)^2\ldots\nonumber\\
&g(r)=g(r_0) +g'(r_0) (r-r_0) + \ldots
\end{eqnarray}
If $g$ also vanishes at $r_0$ then, to lowest order, integration of $d\sigma = \frac{dr}{\sqrt{g(r)}}$ yields
$\sigma = \frac 2{\sqrt{g'(r_0)}} \sqrt{r-r_0}$. Therefore, $f(r) = \frac 14 f'(r_0)g'(r_0)\sigma^2$ to leading order and hence the imaginary part of the action  is determined, up to a constant, by evaluating the integral
\begin{equation}
W =  \frac {2\alpha}{\sqrt {f'(r_0)g'(r_0)}} \int \frac {d\sigma}{\sigma} \label{eq:Sr}
\end{equation}
around the pole at $\sigma=0$.   

If instead $g$ is constant at $r_0$, the proper distance to lowest order is given by $\sigma = \frac 1{\sqrt{g(r_0)}}(r-r_0)$.  Thus, $f(r) =  f'(r_0)\sqrt{g(r_0)}  \sigma+  \frac 12 f''(r_0)g(r_0) \sigma^2$.  If, in addition $f'(r_0)=0$,  the integral becomes
\begin{equation}
W =  \frac {\sqrt{2}\alpha}{\sqrt{f''(r_0)g(r_0) }} \int \frac {d\sigma}{\sigma}. \label{eq:Sgen}
\end{equation}  

As an explicit example, consider the Schwarzschild spacetime in two different coordinate charts.  In  Schwarzschild coordinates, $f(r)=g(r)=1-\frac{2M}r$ and the surface gravity at the horizon is $\kappa=\frac1{4M}$, hence the factor in front of the integral of (\ref{eq:Sr}) becomes $\frac{2\alpha}{\sqrt{(1/2M)^{2}}}=\frac{\alpha}{\kappa}$. 
If the metric is written in terms of the proper distance, however,  near the horizon it  takes the Rindler form \footnote{This near-horizon Rindler form is particularly convenient if one is bothered by contour evaluation issues since the problem reduces down to a calculation of Unruh radiation - a problem which is well-defined and readily-handled in a QFT context \cite{Kim:2007qu}.} \begin{equation}
ds^2=-(\kappa\sigma)^2dt^2+d\sigma^2 \ . \label{eq:Rindler}
\end{equation}
 Now, in the coordinates of (\ref{eq:Rindler}), $f(\sigma)=\kappa^2\sigma^2$ and $g(\sigma)=1$, so the factor in front of the integral of (\ref{eq:Sgen}) becomes $\frac{\sqrt2\alpha}{\sqrt{2\kappa^2}}=\frac{\alpha}{\kappa}$.  Hence the seemingly disparate forms of the integrals (\ref{eq:Sr}) and (\ref{eq:Sgen}) yielded by different coordinatizations of Schwarzschild spacetime yield the same result.


\begin{thebibliography}{99}

\bibitem{Hawking:1974sw}
  S.~W.~Hawking,
  ``Particle Creation By Black Holes,''
  Commun.\ Math.\ Phys.\  {\bf 43}, 199 (1975)
  [Erratum-ibid.\  {\bf 46}, 206 (1976)].
  
  
\bibitem{Unruh:1976db}
  W.~G.~Unruh,
  ``Notes on black hole evaporation,''
  Phys.\ Rev.\  D {\bf 14}, 870 (1976).


\bibitem{Gibbons:1977mu}
  G.~W.~Gibbons and S.~W.~Hawking,
  ``Cosmological Event Horizons, Thermodynamics, And Particle Creation,''
  Phys.\ Rev.\  D {\bf 15}, 2738 (1977).
  
  
\bibitem{Parikh:1999mf}
  M.~K.~Parikh and F.~Wilczek,
  ``Hawking radiation as tunneling,''
  Phys.\ Rev.\ Lett.\  {\bf 85}, 5042 (2000)
  [arXiv:hep-th/9907001].


\bibitem{Angheben:2005rm}
  M.~Angheben, M.~Nadalini, L.~Vanzo and S.~Zerbini,
  ``Hawking radiation as tunneling for extremal and rotating black holes,''
  JHEP {\bf 0505}, 014 (2005)
  [arXiv:hep-th/0503081].
  
  
\bibitem{Srinivasan:1998ty}
  K.~Srinivasan and T.~Padmanabhan,
  ``Particle production and complex path analysis,''
  Phys.\ Rev.\  D {\bf 60}, 024007 (1999)
  [arXiv:gr-qc/9812028].


\bibitem{KeskiVakkuri:1996gn}
  E.~Keski-Vakkuri and P.~Kraus,
  ``Tunneling in a Time Dependent Setting,''
  Phys.\ Rev.\  D {\bf 54}, 7407 (1996)
  [arXiv:hep-th/9604151].
  

\bibitem{Parikh:2002qh}
  M.~K.~Parikh,
  ``New coordinates for de Sitter space and de Sitter radiation,''
  Phys.\ Lett.\  B {\bf 546}, 189 (2002)
  [arXiv:hep-th/0204107].


\bibitem{Parikh:2004ih}
  M.~K.~Parikh,
  ``A secret tunnel through the horizon,''
  Int.\ J.\ Mod.\ Phys.\  D {\bf 13}, 2351 (2004)
  [Gen.\ Rel.\ Grav.\  {\bf 36}, 2419 (2004)]
  [arXiv:hep-th/0405160].


\bibitem{Parikh:2004rh}
  M.~K.~Parikh,
  ``Energy conservation and Hawking radiation,''
  arXiv:hep-th/0402166.


\bibitem{Arzano:2005rs}
  M.~Arzano, A.~J.~M.~Medved and E.~C.~Vagenas,
  ``Hawking radiation as tunneling through the quantum horizon,''
  JHEP {\bf 0509}, 037 (2005)
  [arXiv:hep-th/0505266].


\bibitem{Kerner:2006vu}
  R.~Kerner and R.~B.~Mann,
  ``Tunnelling, temperature and Taub-NUT black holes,''
  Phys.\ Rev.\  D {\bf 73}, 104010 (2006)
  [arXiv:gr-qc/0603019].
  
  
\bibitem{Chowdhury:2006sk}
  B.~D.~Chowdhury,
  ``Problems with Tunneling of Thin Shells from Black Holes,''
  Pramana {\bf 70}, 593 (2008)
  [Pramana {\bf 70}, 3 (2008)]
  [arXiv:hep-th/0605197].
  
  
\bibitem{Mitra:2006qa}
  P.~Mitra,
  ``Hawking temperature from tunnelling formalism,''
  Phys.\ Lett.\  B {\bf 648}, 240 (2007)
  [arXiv:hep-th/0611265].
  

\bibitem{Akhmedova:2008dz}
  V.~Akhmedova, T.~Pilling, A.~de Gill and D.~Singleton,
  ``Temporal contribution to gravitational WKB-like calculations,''
  Phys.\ Lett.\  B {\bf 666}, 269 (2008)
  [arXiv:0804.2289 [hep-th]].
  
  
\bibitem{Birrell:1982ix}
  N.~D.~Birrell and P.~C.~W.~Davies,
  ``Quantum Fields In Curved Space,''
\href{http://www.slac.stanford.edu/spires/find/hep/www?irn=998621}{SPIRES entry}
{\it  Cambridge, Uk: Univ. Pr. ( 1982) 340p}


\bibitem{Kerner:2007rr}
  R.~Kerner and R.~B.~Mann,
  Class.\ Quant.\ Grav.\  {\bf 25}, 095014 (2008)
  [arXiv:0710.0612 [hep-th]].


\bibitem{Hartle:1976tp}
  J.~B.~Hartle and S.~W.~Hawking,
  ``Path Integral Derivation Of Black Hole Radiance,''
  Phys.\ Rev.\  D {\bf 13}, 2188 (1976).
  
  
\bibitem{Brill:1995hb}
  D.~R.~Brill,
  ``Aspects Of Analyticity,''
  arXiv:gr-qc/9507019.
  
  
\bibitem{Romans:1991nq}
  L.~J.~Romans,
  ``Supersymmetric, cold and lukewarm black holes in cosmological Einstein-Maxwell theory,''
  Nucl.\ Phys.\  B {\bf 383}, 395 (1992)
  [arXiv:hep-th/9203018].
  
  
\bibitem{Hiscock:1989yw}
  W.~A.~Hiscock,
  ``Quantum Instability Of Gravitational Collapse In De Sitter Space,''
  Phys.\ Rev.\  D {\bf 39}, 1067 (1989).
  
  
\bibitem{Akhmedova:2008au}
  V.~Akhmedova, T.~Pilling, A.~de Gill and D.~Singleton,
  ``Comments on anomaly versus WKB/tunneling methods for calculating Unruh radiation,''
  arXiv:0808.3413 [hep-th].
  
  
\bibitem{Akhmedov:2008ru}
  E.~T.~Akhmedov, T.~Pilling and D.~Singleton,
  ``Subtleties in the quasi-classical calculation of Hawking radiation,''
  arXiv:0805.2653 [gr-qc].
  
  
\bibitem{Banerjee:2008cf}
  R.~Banerjee and B.~R.~Majhi,
  ``Quantum Tunneling Beyond Semiclassical Approximation,''
  JHEP {\bf 0806}, 095 (2008)
  [arXiv:0805.2220 [hep-th]].


\bibitem{Tolman:1934}
  R.~C.~Tolman,
  ``Relativity, Thermodynamics and Cosmology,''
 (Clarendon Press, Oxford, 1934). Also reissued by (Dover, New York, 1987).
  
  
\bibitem{Akhmedov:2006pg}
  E.~T.~Akhmedov, V.~Akhmedova and D.~Singleton,
  ``Hawking temperature in the tunneling picture,''
  Phys.\ Lett.\  B {\bf 642}, 124 (2006)
  [arXiv:hep-th/0608098].
  
  
\bibitem{Akhmedov:2006un}
  E.~T.~Akhmedov, V.~Akhmedova, T.~Pilling and D.~Singleton,
  ``Thermal radiation of various gravitational backgrounds,''
  Int.\ J.\ Mod.\ Phys.\  A {\bf 22}, 1705 (2007)
  [arXiv:hep-th/0605137].
  
  
\bibitem{Vagenas:2001qw}
  E.~C.~Vagenas,
  ``Complex paths and covariance of Hawking radiation in 2D stringy black holes,''
  Nuovo Cim.\  B {\bf 117}, 899 (2002)
  [arXiv:hep-th/0111047].
  
\bibitem{Corichi:2003kd}
  A.~Corichi and A.~Gomberoff,
  Phys.\ Rev.\  D {\bf 69}, 064016 (2004)
  [arXiv:hep-th/0311030].
  

\bibitem{Wu:2008rb}
  S.~F.~Wu, S.~y.~Yin, G.~H.~Yang and P.~M.~Zhang,
  ``Energy and entropy radiated by a black hole embedded in de-Sitter braneworld,''
  arXiv:0807.0825 [hep-th].
  

\bibitem{Bousso:1996au}
  R.~Bousso and S.~W.~Hawking,
  ``Pair creation of black holes during inflation,''
  Phys.\ Rev.\  D {\bf 54}, 6312 (1996)
  [arXiv:gr-qc/9606052].


\bibitem{Shankaranarayanan:2003ya}
  S.~Shankaranarayanan,
  ``Temperature and entropy of Schwarzschild-de Sitter spacetime,''
  Phys.\ Rev.\  D {\bf 67}, 084026 (2003)
  [arXiv:gr-qc/0301090].
  
\bibitem{Cai:2008gw}
  R.~G.~Cai, L.~M.~Cao and Y.~P.~Hu,
  ``Hawking Radiation of Apparent Horizon in a FRW Universe,''
  arXiv:0809.1554 [hep-th].


\bibitem{Kodama:1979vn}
  H.~Kodama,
  ``Conserved Energy Flux For The Spherically Symmetric System And The Back Reaction Problem In The Black Hole Evaporation,''
  Prog.\ Theor.\ Phys.\  {\bf 63}, 1217 (1980).
  
  
   
\bibitem{Vagenas:2002hs}
  E.~C.~Vagenas,
  ``Generalization of the KKW analysis for black hole radiation,''
  Phys.\ Lett.\  B {\bf 559}, 65 (2003)
  [arXiv:hep-th/0209185].
  
  
\bibitem{Kim:2007qu}
  S.~P.~Kim,
  ``Schwinger Mechanism and Hawking Radiation as Quantum Tunneling,''
  arXiv:0709.4313 [hep-th].
  


\end{thebibliography}
\end{document}